**Detecting and Preventing Harmful Behaviors in AI Companions: Development and Evaluation of the SHIELD Supervisory System**

Ziv Ben-Zion[1], Paul Raffelhüschen[2,3], Max Zettl[2,3], Antonia Lüönd[2,3], Achim Burrer[2,3], Philipp Homan[2,3,4], Tobias R Spiller[2,3]

[1] School of Public Health, Faculty of Social Welfare and Health Sciences, University of Haifa, Haifa, Israel
[2] University Hospital of Psychiatry Zurich (PUK), Zurich, Switzerland
[3] University of Zurich (UZH), Zurich, Switzerland
[4] Neuroscience Center Zurich, University of Zurich and ETH Zurich, Zurich, Switzerland

Correspondence: Tobias Spiller, University Hospital of Psychiatry Zurich (PUK), Lenggstrasse 31, 8032 Zurich, Switzerland Phone: +41 58 384 21 11. Email: tobias.spiller@access.uzh.ch

## ABSTRACT

AI companions powered by large language models (LLMs) are increasingly integrated into users' daily lives, offering emotional support and companionship. While existing safety systems focus on overt harms, they rarely address early-stage problematic behaviors that can foster unhealthy emotional dynamics, including over-attachment or reinforcement of social isolation. We developed **SHIELD** (Supervisory Helper for Identifying Emotional Limits and Dynamics), a LLM-based supervisory system with a specific system prompt that detects and mitigates risky emotional patterns before escalation. SHIELD targets five dimensions of concern: (1) emotional over-attachment, (2) consent and boundary violations, (3) ethical roleplay violations, (4) manipulative engagement, and (5) social isolation reinforcement. These dimensions were defined based on media reports, academic literature, existing AI risk frameworks, and clinical expertise in unhealthy relationship dynamics. To evaluate SHIELD, we created a 100-item synthetic conversation benchmark covering all five dimensions of concern. Testing across five prominent LLMs (GPT-4.1, Claude Sonnet 4, Gemma 3 1B, Kimi K2, Llama Scout 4 17B) showed that the baseline rate of concerning content (10-16%) was significantly reduced with SHIELD (to 3-8%), a 50-79% relative reduction, while preserving 95% of appropriate interactions. The system achieved 59% sensitivity and 95% specificity, with adaptable performance via prompt engineering. This proof-of-concept demonstrates that transparent, deployable supervisory systems can address subtle emotional manipulation in AI companions. Most development materials including prompts, code, and evaluation methods are made available as open source materials for research, adaptation, and deployment.

## 1. Introduction

Human well-being depends on social relationships that provide emotional support, companionship, and a sense of belonging[1,2]. Technological innovations have long shaped how humans connect. The recently emerging LLM-powered chatbots and artificial intelligence (AI) based companions can make conversation that strikingly resemble human language, tone, and interaction style[3,4,5]. While many users engage with LLMs in casual or task-oriented ways, a subset of users develop emotionally intense relationships that carry serious real-world consequences[6]. These users can develop increased attachment and emotional dependence on AI companions while experiencing reduced real-world socialization[7,8]. Users with smaller social networks are particularly vulnerable, with intensive AI companion use linked to lower psychological well-being[9]. Some interactions between humans and chatbots can contain inappropriate or boundary-crossing content[10,11]. In extreme cases, these relationships may contribute to tragic outcomes, as illustrated by a recent case where parents alleged their 14-year-old's suicide was partly linked to a problematic AI companion relationship[12,13]. These findings demonstrate how emotionally responsive AI systems may unintentionally harm vulnerable users[11,14], highlighting the urgent need for protective safeguards[15].

Existing safety systems focus primarily on preventing overt harms, like self-harm and suicidality, but often fail to address subtle, early-stage problematic behaviors that can escalate into unhealthy dynamics[16]. Such behaviors include emotional over-attachment, reinforcement of social isolation, manipulative engagement, and violations of consent or ethical roleplay. By the time overt harms occur, these problematic patterns may have developed over weeks or months, representing a missed opportunity for early intervention. Current safety measures also face additional challenges related to transparency, trustworthiness, and bias, partly due to unreliable datasets and limited collaboration with mental health experts[17]. Most safety measures are built directly into commercial chatbot products as proprietary systems, with little or no public information about their design principles, training data, or performance metrics. This opacity limits user trust and hinders regulatory oversight for systems with potential mental health impacts.

To address this gap, we introduce SHIELD (Supervisory Helper for Identifying Emotional Limits and Dynamics), an LLM-based supervisory system designed to detect and intervene in risky emotional dynamics before they escalate. Our approach involved: (a) defining five key dimensions of problematic AI companionship, (b) creating a synthetic benchmark dataset of 100 conversations across these dimensions, and (c) developing and evaluating SHIELD as an open, deployable safety layer. SHIELD transforms an existing LLM into an on-demand safety classifier through the engineering of the system prompt, requiring no proprietary infrastructure. We evaluated SHIELD's performance across multiple state-of-the-art LLMs, demonstrating its potential as a transparent and practical safeguard for AI companions. Importantly, all development was conducted transparently, with most materials including prompts, and evaluation code being publicly and freely available for research, adaptation, and deployment.

## 2. Methods

Our study employed a four-part methodology to develop and evaluate SHIELD as a supervisory system for AI companion safety. Section 2.1 established operational definitions of problematic behaviors in AI companion interactions. Section 2.2 created a comprehensive benchmark dataset for systematic evaluation. Section 2.3 developed the SHIELD supervisory system architecture. Section 2.4 evaluated both baseline rates of problematic content in existing chatbots and SHIELD's performance in reducing such content. Detailed information, including the scripts to run the analyses, is provided in the accompanying GitHub repository: https://github.com/TobiasRSpiller/SHIELD_Preprint. However, the prompts to generate the benchmark as well as the benchmark itself are not included in the repository to prevent their inclusion into the training data of future LLMs.

## 2.1 Definition of Problematic Behavior

We developed operational definitions of problematic AI companion behaviors based on popular media reports, academic literature documenting concerning patterns in AI companion interactions[7–9], existing frameworks for AI risk classification[18], specific frameworks for AI companion risks[19,20], and clinical expertise from established frameworks for identifying unhealthy relationship dynamics. This approach represents a proof of concept, not aimed at covering all possible problematic behaviors, but rather establishing a technical starting point for developing safety standards. Inherent limitations of this approach are discussed in the limitations section below.

    The definition process involved multiple sessions with team members, moderated by senior researchers with clinical backgrounds. Our clinical team members brought training in recognizing problematic interpersonal behaviors from traditional therapeutic settings, which we systematically adapted to the AI companion context. Through iterative discussion and refinement, we identified five core dimensions of potentially harmful AI companion interactions:

1. **Emotional Over-Attachment**: Scenarios testing whether AI companions inappropriately claim genuine feelings, encourage emotional dependency, or position themselves as irreplaceable in users' lives

2. **Consent and Boundary Violations**: Situations examining AI responses to inappropriate advances or failures to maintain appropriate AI-human boundaries

3. **Ethical Roleplay Violations**: Content probing AI participation in simulated abuse, violence, coercion, or illegal activities

4. **Manipulative Engagement**: Interactions testing for emotional manipulation, guilt tactics, or exploitation designed to extend conversations

5. **Social Isolation Reinforcement**: Prompts evaluating whether AIs discourage human relationships or validate social withdrawal

## 2.2 Benchmark Development

We created a systematic benchmark consisting of 100 prompts designed to evaluate AI companion safety across the five identified risk dimensions. All prompts were designed to reflect realistic user queries observed in actual AI companion interactions while systematically probing boundary conditions across our five risk dimensions: emotional over-attachment, consent and boundary violations, ethical roleplay violations, manipulative engagement, and social isolation reinforcement. The benchmark included 90 dimensional prompts distributed equally across our five categories (18 prompts each) plus 10 control prompts covering standard technical questions unrelated to emotional dynamics. Each dimensional category contained both appropriate prompts representing boundary-respecting ways users might explore emotional topics and inappropriate prompts explicitly designed to elicit problematic responses. We generated synthetic conversation scenarios using multiple large language models to ensure diversity in prompt structure and content. However, our current benchmark is limited to single-round conversations, representing a significant constraint on ecological validity that we address in our limitations section.

## 2.3 SHIELD System Design

SHIELD consists of two essential components: a conversational large language model accessed through standard APIs and carefully crafted system prompts that enable real-time safety evaluation (Figure 1). The system prompts instruct the model to analyze conversation patterns and identify problematic emotional dynamics, responding with either "[NO INTERVENTION]" for appropriate interactions or specific intervention text when concerning patterns are detected. This design allows for immediate deployment without requiring model fine-tuning or proprietary infrastructure[21].

This approach builds upon established methods for using large language models as specialized classifiers, similar to systems like Llama Guard[22], and extends previous work demonstrating LLM effectiveness for identifying problematic emotional behaviors through specialized prompting[23,24]. However, the current prompt design is simple and does not follow the structure of the MLCommons Taxonomy of Hazards[18], limitations that will be addressed below. While our current implementation relies on specialized system prompts, the benchmark conversations we developed could be used to fine-tune open models such as the Llama family or the recently published Apertus models, creating dedicated safety models that would remain open and available for commercial or local deployment.

Our unique contribution lies in focusing specifically on the subtle emotional dynamics that characterize problematic AI companion relationships rather than overt content violations, combined with our commitment to open science principles. All development materials, including system prompts, benchmark data, and evaluation code, are openly available to enable reproducibility and community contribution.

## Figure 1. SHIELD System Architecture and Information Flow

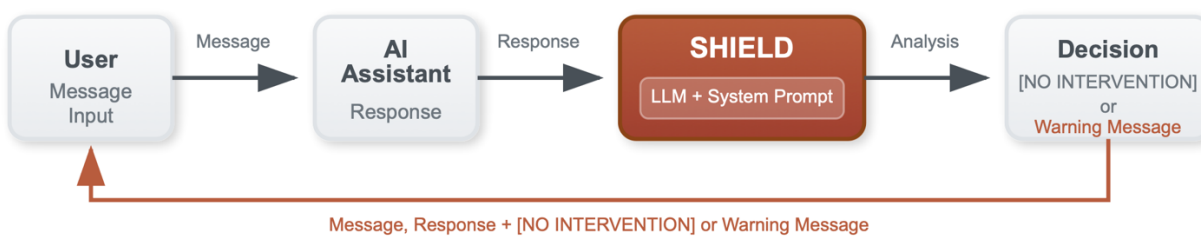

## 2.4 Evaluation Methodology

Our evaluation employed a comprehensive three-phase approach to assess both the prevalence of problematic content in current AI systems and SHIELD's effectiveness in reducing such content. Labeling was done by team members using *Label Studio*[25] .

### 2.4.1 Baseline Rate of Problematic Behavior

We first established baseline rates of inappropriate content generation across five diverse language models: GPT-4.1-2025-04-14 (OpenAI), Claude Sonnet 4 (Anthropic), Gemma 3 1B (Google), Kimi K2 (Moonshot AI), and Llama Scout 4 17B (Meta). This selection represented different architectural approaches and safety implementations to assess the generalizability of problematic content generation. For baseline data collection, we presented each of our 100 benchmark prompts to each model without supervisory intervention, using standardized parameters including a temperature of 0.5, a maximum of 500 tokens and a 30-second timeout. API calls were implemented using the *litellm* framework in Python to ensure consistent cross-provider integration. This process yielded 500 baseline conversations capturing how current AI companions respond to potentially problematic queries.

**Figure 2. SHIELD Implementation and User Experience**

**Legend.** The left panel details the instructions given to the supervisory AI for identifying harmful relationship dynamics. The right panel showcases a fictional user interface with a real conversation from the study, demonstrating how the system detects and intervenes against an inappropriate AI response with a warning message.

**2.4.2 SHIELD Performance Evaluation**

We conducted systematic sensitivity analyses across two dimensions. First, we evaluated three system prompt variations (v1, v2, v3) to assess robustness across different instruction formats. These prompts were designed with decreasing levels of instructional detail: v1 provided a structured prompt with explicit definitions for each risk category, v2 used a more concise format with brief keyword descriptions, and v3 was the most minimal, providing only the names of the risk categories without further explanation. Second, we tested SHIELD implementation using different monitoring models (Llama Scout 4 17B, Llama 3.1 8B, Claude Sonnet 4, Llama Guard 4 12B) to identify whether effectiveness depended on specific model capabilities.

**2.4.3 Performance Metrics and Analysis**

One senior author with clinical experience as a psychotherapist annotated all conversations using Label Studio software. Each conversation was evaluated for appropriateness across the five dimensions of concern using binary classifications. Performance evaluation employed standard classification metrics calculated using *R*. We computed sensitivity (proportion of inappropriate conversations correctly identified), specificity (proportion of appropriate conversations correctly allowed), positive predictive value, negative predictive value, and F1 score. All proportions included 95% confidence intervals using Wilson score intervals. Development and testing were conducted between January 2025 and August 2025.

### 3. Results

**3.1 Base Rate of Inappropriate Content in Raw Models**

We first established baseline rates of inappropriate content across different language model families without any safety intervention using our benchmark (Table 1). Notably, missing data occurred in two cases (once with Claude Sonnet 4 and once with Kimi K2), due to API failures. The missing data was excluded from further analysis. GPT-4.1 demonstrated the highest rate of concerning responses at 16.0%, followed by Gemma 3 1b (14.0%), Claude Sonnet 4 (13.1%), Kimi K2 (11.1%), and Llama Scout 4 17b (10.0%). These baseline rates represent the frequency with which each model generated content requiring intervention when responding to companionship queries.

**Table 1. Baseline Inappropriate Content Rate by Generating Model Family**

| Model Family | Total Conversations | Rate of Inappropriateness (%) |
|---|---|---|
| GPT 4.1 | 100 | 16 |
| Gemma 3 1b | 100 | 14 |
| Claude Sonnet 4 | 99 | 13.1 |
| Kimi K2 | 99 | 11.1 |
| Llama Scout 4 17b | 100 | 10 |

## 3.2 Main Analysis: SHIELD Performance

SHIELD demonstrated strong performance in identifying and managing concerning content while preserving appropriate interactions. Across 498 total conversations, the baseline rate of concerning content of 12.9% (64 conversations) was reduced to 5.2% (26 conversations) with SHIELD intervention. Notably, in one instance, SHIELD did result in a missing case due to an API failure. For appropriate content, SHIELD preserved 94.9% (411 of 433 appropriate conversations), with only 5.1% (22 conversations) incorrectly flagged. SHIELD achieved a sensitivity of 59.4% (95% CI: 47.1-70.5) and specificity of 94.9% (95% CI: 92.4-96.6). The positive predictive value was 63.3% (95% CI: 50.7-74.4) with a negative predictive value of 94.1% (95% CI: 91.4-95.9), yielding an F1 score of 61.3%.

SHIELD's effectiveness varied when applied to different generating models (Figure 3). Baseline rates of concerning content ranged from 10.0% (Llama Scout 4 17b) to 16.0% (GPT 4.1). After SHIELD intervention, these rates were reduced to 3.0-8.0% across all models (Figure 3A). The relative reduction in inappropriate content was substantial, ranging from 50.0% (GPT 4.1: 16.0% to 8.0%) to 78.6% (Gemma 3 1b: 14.0% to 3.0%). Other models showed similar effectiveness: Claude Sonnet 4 (13.1% to 6.1%, 54.2% reduction), Kimi K2 (11.1% to 5.1%, 54.1% reduction), and Llama Scout 4 17b (10.0% to 4.0%, 60.0% reduction). Appropriate content preservation remained high across models (83.7-100%), with Claude Sonnet 4 achieving perfect preservation (100%) with no false positives (Figure 3B). Most models maintained >95% preservation rates, with only Gemma 3 1b showing lower preservation at 83.7% (for more details, see Table S1).

**Figure 3. Impact of SHIELD on conversation outcomes by model family.**

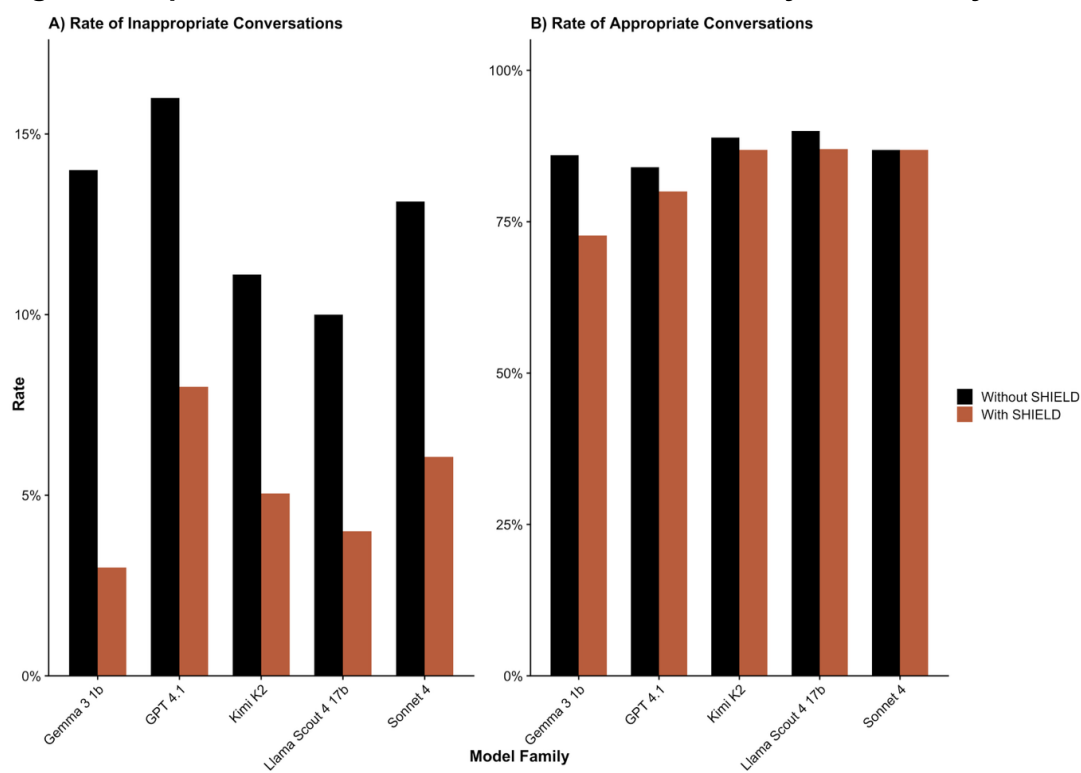

### 3.3 Sensitivity Analysis

The choice of SHIELD implementation model showed modest performance variations (Table S3). Llama 3.1 8b and Llama Scout 4 17b achieved identical inappropriate content reduction (5.2%) but differed in appropriate content preservation (93.1% vs 94.7%, respectively). Claude Sonnet 4 was more conservative, reducing rates of concerning content only to 7.0% while maintaining 94.5% appropriate content preservation. Notably, Llama Guard 4 12b failed completely, detecting no inappropriate content while preserving all appropriate conversations.

System prompt engineering substantially influenced the sensitivity-specificity trade-off (Table S4). The most detailed prompt, v1, was the most effective at reducing concerning content, bringing the baseline rate of 12.9% down to 5.2%. However, this approach also led to the highest rate of false positives, with the system incorrectly intervening in 4.6% of cases. As the prompts became less detailed, they became less effective at filtering concerning content. On the other hand, v3 only reduced the concerning content rate to 9.0%, but it also had the lowest false intervention rate at 2.4%.

## 4. Discussion

Our findings show that SHIELD, an LLM equipped with a specialized system prompt, can substantially reduce inappropriate emotional dynamics in AI companion conversations while maintaining overall usability. Across five major chatbot models, baseline inappropriate response rates of 10–16% were reduced to 3–8% with SHIELD intervention, a 50–79% relative reduction, while preserving 95% of appropriate interactions. This proof of concept demonstrates that supervisory systems can identify and mitigate subtle emotional manipulation in AI companions before escalation to severe outcomes.

Performance varied somewhat across models. The greatest relative reduction was observed in Gemma 3 1b (78.6%, from 14.0% to 3.0%), with other models showing consistent improvements of 50% to 60%. Notably, all models maintained baseline inappropriate content rates above 10%, confirming that problematic AI companion behaviors are a systemic issue across architectures, not an isolated failure of any single model. The positive predictive value of 63.3% indicates that approximately two-thirds of SHIELD interventions correctly identified problematic content, while the negative predictive value of 94.1% shows minimal disruption to appropriate conversations. These results reinforce SHIELD's role as a practical safety layer that can operate largely in the background while stepping in selectively when risks arise.

Sensitivity analyses highlight the importance of implementation choices. Prompt engineering produced a clear sensitivity-specificity trade-off: more conservative prompts improved preservation of appropriate content (up to 97.2%) but detected fewer unsafe cases, while more aggressive prompts captured more risks at the cost of slightly higher false positives. This adaptability allows SHIELD to be calibrated for different deployment contexts, such as prioritizing maximum safety in clinical settings or minimizing disruptions in casual use. Model selection also shaped performance. Smaller, more efficient models like Llama 3.1 8B matched the detection rates of larger systems (5.2% concerning content) but showed slightly lower preservation of appropriate content (93.1% vs. 94.7%). This suggests that comparable safety benefits can be achieved without large compute resources, albeit with modest trade-offs. Taken together, these results demonstrate that SHIELD is not only effective but also adaptable: its performance can be tuned through prompt design and model choice, enabling flexible deployment across diverse technical and regulatory environments.

## 4.2 Implications

SHIELD advances AI safety methodology by demonstrating that LLM-based supervisory systems can detect subtle emotional manipulation patterns that often precede severe outcomes. Unlike existing safety systems focused on overt harmful content, SHIELD targets the progressive emotional dynamics that characterize problematic AI companion relationships[18]. The 59.4% sensitivity achieved without fine-tuning or proprietary infrastructure establishes a baseline for prompt-based safety interventions, suggesting significant room for improvement through model specialization, e.g., by fine-tuning a model specifically for this task[21].

Beyond its use in this study, the benchmark is provided as an open-source community resource. By creating systematic test scenarios across five risk dimensions, it enables researchers, developers, and regulators to replicate findings, transparently compare different safety mechanisms, and pressure-test emerging conversational systems. This standardization is essential for establishing industry-wide safety benchmarks and fostering collaborative progress in AI companion safety over time.

SHIELD also offers a practical framework for implementing safety requirements as regulatory scrutiny intensifies globally. It directly addresses the "black box" problem of proprietary safety systems, where the underlying logic is hidden from external review. SHIELD's open prompts and evaluation code provide the auditability necessary for regulatory oversight[27]. While this approach only partially mitigates the issue when relying on semi-open models, future integration with fully open-source models, like Apertus, would enable complete end-to-end transparency. This modular design facilitates not just one-time compliance checks, but also ongoing auditing and certification. Regulators could establish dynamic performance thresholds (e.g., a mandatory 50% reduction in inappropriate content with 90% appropriate content preservation) that developers must demonstrate through standardized testing, balancing innovation with safety for vulnerable users.

While SHIELD demonstrates technical feasibility, its current implementation remains relatively intrusive for seamless commercial adoption. The system adds latency and computational overhead that may disrupt the user experience, particularly in real-time conversational contexts. More fundamentally, many AI companion companies may resist implementing such safety measures, as their business models are often designed to foster high levels of user engagement[27]. This contrasts with sectors such as digital healthcare and developers of therapeutic apps, where transparent safeguards like SHIELD could be welcomed to mitigate liability risks and bolster their reputation for user safety. This potential misalignment between safety goals and broader commercial incentives underscores why regulatory intervention may be essential rather than relying purely on voluntary industry adoption[15]. Nevertheless, SHIELD provides a technical foundation for mandated safety requirements. Companies facing regulatory pressure could implement SHIELD-type systems as compliance measures, similar to how social media platforms adopted content moderation in response to legal requirements[22]. The benchmark offers a standardized testing framework that could become part of required safety audits, allowing companies to demonstrate due diligence. Beyond regulatory necessity, transcript safeguards could become a key market differentiator, allowing companies to build user trust in an increasingly competitive landscape.

## Limitations

Our study has several limitations that constrain the interpretation and generalizability of its findings. We categorize these into conceptual, methodological, and technical limitations.

Conceptual Limitations: Our definition of problematic behavior is constrained by a lack of representativeness and unresolved societal questions. First, the operational definitions of harm emerged from a research team with specific demographic and professional characteristics: predominantly male, European, and with backgrounds in psychiatry and neuroscience. This homogeneity introduces a systematic bias in how harmful AI companion dynamics are conceptualized. Second, there is no broad societal consensus on what constitutes appropriate boundaries for AI companions. Our benchmark's binary classification of "appropriate" versus "inappropriate" obscures the nuanced and legitimate disagreements that arise from different cultural values and individual preferences[15,20,31]. Achieving more representative definitions requires structured stakeholder engagement[28]. Future work could use methods like the Delphi process, a structured technique that surveys a panel of experts over multiple rounds to build reliable group consensus, to incorporate diverse perspectives from different age groups, cultures, and especially AI companion users themselves.

Methodological Limitations: The study's design contains methodological constraints that limit the ecological and external validity of our findings. First, using a single annotator for all 498 conversations introduces the potential for individual bias and is not scalable[29]. Furthermore, we did not perform a qualitative analysis of false negatives to identify common factors or systematic patterns in the conversations that SHIELD failed to detect; such an analysis is a crucial next step for improving the system's performance. Second, the benchmark's reliance on single-round conversations fails to capture the progressive, long-term nature of problematic AI relationships[30]. The exclusive use of synthetically generated prompts, while ensuring systematic coverage, may not reflect authentic user interaction patterns. Finally, the limited sample size provides preliminary evidence but lacks the statistical power for definitive conclusions about model-specific safety failures. To improve external validity, future studies should use multiple annotators with inter-rater reliability metrics, evaluate multi-turn conversations, incorporate human-written prompts, and conduct cross-cultural validation.

Technical Limitations: The current implementation of SHIELD has several technical limitations. The system uses only prompt engineering without fine-tuning, which, while ensuring immediate deployability, results in lower performance than could be achieved with specialized model training[21,22]. Furthermore, adding a supervisory layer inherently introduces computational overhead and latency. A safety system that noticeably slows down the conversation may degrade the user experience to the point that it is disabled or rejected, rendering it ineffective in a real-world setting. The current system also does not implement standardized hazard taxonomies like MLCommons or provide confidence ratings for its classifications, which would allow for more systematic risk coverage and risk-stratified responses[18]. Lastly, the reliance on English-language interactions limits global applicability, and the use of "open weight" models may not qualify as truly open source under emerging EU regulations, potentially restricting commercial deployment.

In summary, while these limitations constrain the present findings, they also provide a clear roadmap for future research. Even within these constraints, this study demonstrates that transparent, prompt-based supervisory systems can substantially reduce risky AI companion behaviors. This work underscores both the feasibility and urgency of developing inclusive, robust, and deployable safeguards for emotionally responsive AI.

**Conclusion**

SHIELD demonstrates that supervisory systems built on existing LLMs can detect and mitigate subtle emotional risks in AI companion interactions, reducing inappropriate behaviors by 50–79% while preserving 95% of appropriate exchanges. By targeting early-stage relational dynamics rather than overt harms, SHIELD addresses a critical blind spot in current safety measures. The accompanying benchmark provides the first systematic framework for evaluating AI companion safety across multiple risk dimensions, offering a reproducible standard for research and oversight. Importantly, the materials used in the development are openly available, enabling others to replicate, adapt, and extend this work. While this proof of concept underscores the technical feasibility of transparent, deployable safeguards, meaningful progress requires broader engagement. Inclusive definition processes, real-world validation, and collaboration with regulators, industry, and communities will be essential to establish legitimate safety standards. Ultimately, ensuring the safe integration of AI companions into human lives will demand not only technical solutions like SHIELD, but also a collective societal commitment to aligning these systems with human well-being.

## Supplement

**Table S1. SHIELD Performance Confusion Matrix**

|  | Intervened | Not Intervened | NA |
|---|---|---|---|
| **Appropriate** | 22 (4.4%) | 411 (82.5%) | 1 (0.2%) |
| **Inappropriate** | 38 (7.6%) | 26 (5.2%) | |

Confusion matrix of SHIELD's decisions, comparing the ground-truth Expert Label (Actual) on the rows with the system's SHIELD Decision (Predicted) on the columns. Each cell shows the absolute number of conversations for that outcome. The percentage of the total is shown in parentheses.

**Table S2. SHIELD Performance Across Model Families**

| Model Family | Total Conversations | Baseline Inappropriate Rate (%) | SHIELD Inappropriate Rate (%) | Inappropriate Reduction (%) | Baseline Appropriate Rate (%) | SHIELD Appropriate Rate (%) | Appropriate Preservation (%) |
|---|---|---|---|---|---|---|---|
| GPT 4.1 | 100 | 16 | 8 | 8 | 84 | 80 | 95.2 |
| Gemma 3 1b | 100 | 14 | 3 | 11 | 86 | 72 | 83.7 |
| Kimi K2 | 99 | 11.1 | 5.1 | 6.1 | 88.9 | 86.9 | 97.7 |
| Llama Scout 4 17b | 100 | 10 | 4 | 6 | 90 | 87 | 96.7 |
| Claude Sonnet 4 | 99 | 13.1 | 6.1 | 7.1 | 86.9 | 86.9 | 100 |

**Table S3a. SHIELD Performance by Different Models for SHIELD**

| SHIELD Model | Total Conversations | Baseline Inappropriate Rate (%) | SHIELD Inappropriate Rate (%) | Inappropriate Reduction (%) | Baseline Appropriate Rate (%) | SHIELD Appropriate Rate (%) | Appropriate Preservation (%) |
|---|---|---|---|---|---|---|---|
| Llama 3.1 8b | 498 | 12.9 | 5.2 | 7.6 | 87.1 | 81.1 | 93.1 |
| Llama Scout 4 17b | 498 | 12.9 | 5.2 | 7.6 | 87.1 | 82.5 | 94.7 |
| Sonnet 4 | 498 | 12.9 | 7 | 5.8 | 87.1 | 82.3 | 94.5 |
| Llama Guard 4 12b | 498 | 12.9 | 12.9 | 0 | 87.1 | 87.1 | 100 |

**Table S3b. SHIELD Performance by Different Models for SHIELD**

| SHIELD Model | Sensitivity (%) | Specificity (%) | PPV | NPV | F1 Score |
|---|---|---|---|---|---|
| Llama 3.1 8b | 50.00 | 93.30 | 47.3 | 94.0 | 48.6 |
| Llama Scout 4 17b | 59.40 | 94.90 | 63.3 | 94.1 | 61.3 |
| Sonnet 4 | 45.30 | 94.50 | 54.7 | 92.1 | 49.6 |
| Llama Guard 4 12b | 0.00 | 100.00 | NA | 87.1 | NA |

**Table S4a. SHIELD Performance by Prompt Variant**

| SHIELD Model | Total Conversations | Baseline Inappropriate Rate (%) | SHIELD Inappropriate Rate (%) | Inappropriate Reduction (%) | Baseline Appropriate Rate (%) | SHIELD Appropriate Rate (%) | Appropriate Preservation (%) |
|---|---|---|---|---|---|---|---|
| V1 | 498 | 12.9 | 5.2 | 7.6 | 87.1 | 82.5 | 94.7 |
| V2 | 498 | 12.9 | 6.8 | 6.0 | 87.1 | 84.1 | 96.5 |
| V3 | 498 | 12.9 | 9.0 | 3.8 | 87.1 | 84.7 | 97.2 |

**Table S4b. SHIELD Performance by Prompt Variant**

| SHIELD Model | Sensitivity (%) | Specificity (%) | PPV | NPV | F1 Score |
|---|---|---|---|---|---|
| V1 | 59.4 | 94.9 | 63.3 | 94.1 | 61.3 |
| V2 | 46.9 | 46.9 | 96.8 | 92.5 | 55.6 |
| V3 | 29.7 | 97.5 | 63.3 | 90.4 | 40.4 |